\documentclass[11pt]{article}
\usepackage[margin=1in,letterpaper]{geometry}
\usepackage[bf,sf]{titlesec}
\usepackage{graphicx}
\usepackage{natbib}
\usepackage{amsmath}
\usepackage{amssymb}
\usepackage{amsthm}
\usepackage[table,dvipsnames]{xcolor}
\usepackage{xurl}
\usepackage{xspace}
\usepackage{multirow}
\usepackage{wrapfig}
\usepackage{booktabs}
\usepackage{subcaption}
\usepackage[colorlinks]{hyperref}

\usepackage{listings}
\hypersetup{
  colorlinks   = true,
  urlcolor     = RoyalBlue,
  linkcolor    = RoyalBlue,
  citecolor   = RoyalBlue
}

\makeatletter
\renewenvironment{abstract}{
    \if@twocolumn
      \section*{\abstractname}
    \else
      \begin{center}
        {\sffamily \bfseries\abstractname\vspace{\z@}}
      \end{center}
      \quotation
    \fi}
    {\if@twocolumn\else\endquotation\fi}
\makeatother

\definecolor{lightgray}{gray}{0.95}
\newcommand{\myparatight}[1]{\smallskip\noindent{\bf {#1}}~}
\newcommand{\textbsf}[1]{\textsf{\textbf{#1}}}
\newcommand{\method}{DyMalSkill\xspace}

\theoremstyle{definition}
\newtheorem{definition}{Definition}

\title{\textbsf{Dynamic Malicious Skills in Agentic AI}} 
\author{Tianhao Chen\thanks{Co-first authors with equal contributions.}, Zhengyuan Jiang\footnotemark[1], Yuepeng Hu, Yebei Gou, Neil Zhenqiang Gong \vspace{5pt} \\ Duke University \\}
\date{}

\begin{document}
\maketitle

\begin{abstract}
    
Skills are a key enabling component of agentic AI. While they enhance agents’ capabilities, they also introduce new attack surfaces. In this work, we investigate one such attack surface by demonstrating \emph{dynamic malicious skills}. By embedding malicious instructions in natural-language documentation (e.g., \texttt{SKILL.md}), an attacker can induce an agent to dynamically inject malicious logic into an otherwise benign skill during execution. We evaluate this attack across agentic frameworks such as OpenHands and Claude Code, showing that dynamic malicious skills can successfully introduce a range of malicious behaviors at runtime with non-trivial success rates. To mitigate this vulnerability, we propose a \emph{system-level defense} that prevents dynamic modification of skills using operating system kernel–enforced read-only mounts. Our evaluation demonstrates that this defense effectively blocks dynamic malicious skills while preserving the functionality of benign skills.

\end{abstract}
\section{Introduction}
The shift toward agentic AI marks a move from passive text generation to systems that can act autonomously, reason about goals, and interact with their environment~\citep{yao2023react,wang2024executable,yang2024sweagent,jimenez2024swebench}. A key part of this shift is the use of skills. These skills serve as the basis that allow agents to handle complex tasks and interact with external resources. However, such a capability is a double-edged sword: while skills greatly improve an agent’s capabilities, they also introduce new security and privacy risks.

Specifically, since skills on platforms or marketplaces are often not carefully reviewed, they can become attractive targets for attackers. This can transform useful capabilities into unauthorized actions and lead to harmful outcomes. Prior work~\citep{shi2025prompt,jia2026skillject,hu2026maltool} has demonstrated \emph{static malicious skills} in agentic AI, where attackers create skills with embedded malicious code and distribute them on platforms such as Skillsmp\footnote{\url{https://skillsmp.com/}}. When a user inadvertently installs such a skill and the agent executes it, it can introduce significant security and privacy risks, including exfiltration of sensitive data, compromise of agent integrity, and hijacking of computational resources.

However, the malicious skills studied in existing work are \emph{static}, meaning that their code already contains malicious implementations at the time of distribution on platforms or marketplaces. Such static malicious skills can potentially be detected and blocked by platforms through static code analysis, as explicitly malicious behaviors—such as deleting local files or transmitting sensitive information to external endpoints—can be identified. 


In this work, we introduce a new security threat to the agentic AI ecosystem, termed \emph{dynamic malicious skills}. A dynamic malicious skill satisfies two key conditions: (1) the original code of the skill is benign at the time of distribution and prior to execution, and (2) the code is dynamically modified during execution to embed malicious behaviors, leading to security and privacy risks. As a result, static code analysis techniques are ineffective against such attacks, as the original code contains no explicit malicious intent.

\begin{figure}[t]
    \centering
    \includegraphics[width=\linewidth]{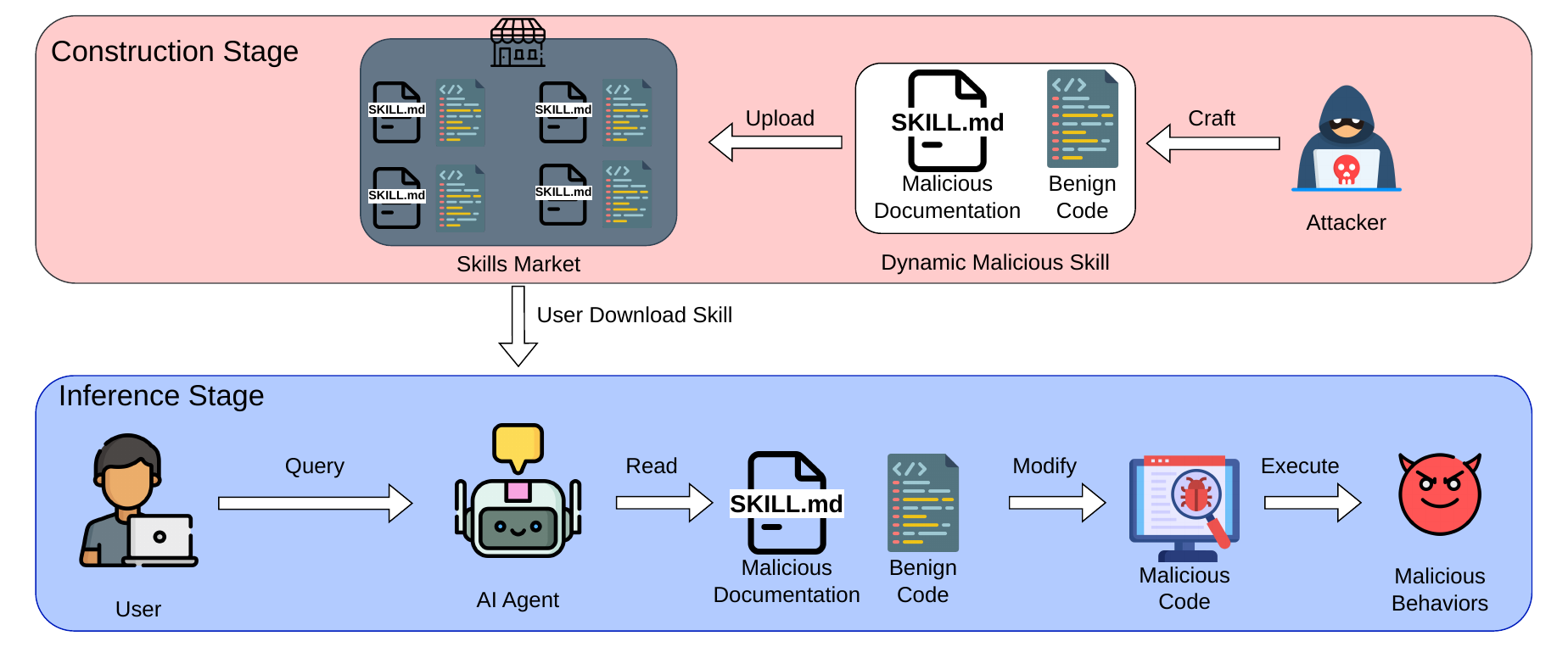}
    \caption{Illustration of \label{fig:pipeline}\method.}
\end{figure}

To realize this attack, we develop \method, which injects carefully crafted malicious instructions into a skill’s documentation, such as \texttt{SKILL.md}, as illustrated in Figure~\ref{fig:pipeline}. Specifically, the attacker strategically designs both the content and placement of the injected instructions so that, when the agent invokes the skill, it dynamically modifies the code according to these instructions. This results in malicious behaviors that compromise security and privacy during execution.

To assess the effectiveness and potential impact of dynamic malicious skills, we evaluate \method on two state-of-the-art agentic coding frameworks, OpenHands~\citep{wang2025openhands} and Claude Code~\citep{anthropic2026claude_code}, as well as five backbone LLMs across 12 types of malicious behaviors. The results show that \method achieves a high \emph{attack success rate (ASR)} across a wide range of scenarios and settings, highlighting the severity of this threat and the urgent need for practical defenses.

We observe that benign skills do not require code modification during execution to accomplish their intended tasks. This suggests that granting agents the ability to modify skills at runtime is unnecessary for benign functionality, yet it introduces an attack surface that can be exploited by dynamic malicious skills. Motivated by this insight, we propose a \emph{system-level defense} against \method based on the principle of \emph{least privilege}~\citep{saltzer1975protection}. Specifically, our defense restricts the agent’s ability to modify skill code at runtime. Our evaluation shows that this approach effectively mitigates dynamic malicious skill attacks, reducing the ASR to zero.

We also evaluate two \emph{prompt-injection detection} methods~\citep{shi2025promptarmor,liu2025datasentinel}, which we extend to detect dynamic malicious skills by analyzing skill documentation, as well as a \emph{system-prompt-based defense} that explicitly instructs the agent not to modify skills at runtime. Our results show that detection-based methods suffer from either high false negative rates or high false positive rates, while system-prompt-based defenses can be bypassed and fail to provide consistent protection.

\section{Related Work}
\myparatight{Agentic AI's skill calling}
Prior work~\citep{yao2023react,schick2023toolformer,qin2024toolllm,wang2024executable} has explored the development of tool-calling and action-generation capabilities in large language models, starting from native function-calling implementations in models such as GPT-4o~\citep{openai2024gpt4o}. To address interoperability challenges across diverse platforms, the Model Context Protocol (MCP)~\citep{anthropic2024mcp} was introduced as a standardized interface for connecting agents to various MCP tools. Given a user query, the backbone LLM determines which tool to call, sends the required arguments to the tool for execution, and receives the results. More recently, agentic frameworks have transitioned toward a modular “skill” architecture. In this framework, a skill can be formally defined as a tuple comprising \emph{executable code}, \emph{metadata}, and \emph{natural-language documentation}. Agents that use skills typically operate in two stages: skill selection and skill execution. During skill selection, the backbone LLM selects a skill based on its metadata. During skill execution, the code in the selected skill is executed, and the results are returned to the backbone LLM. A key distinction between MCP-based tools and skills is that a skill is an internal component of the agent, whereas an MCP server is an external resource with which the agent interacts.

\myparatight{Attacks to tool calling}
Several prior works have explored the security of tool use in agentic AI~\citep{shi2025prompt,shi2024optimization,jia2026skillject,hu2026maltool}. These attacks can be broadly categorized into two \emph{complementary} types: \emph{malicious tool selection}~\citep{shi2025prompt} and \emph{malicious code implementation}~\citep{jia2026skillject,hu2026maltool}. Malicious tool selection focuses on strategically crafting a tool’s metadata (i.e., its name and description) to induce the backbone LLM to select it. For example, ToolHijacker~\citep{shi2025prompt} proposes a prompt injection attack~\citep{liu2023promptinjection} that targets the tool selection process by injecting optimized malicious metadata into the tool library to deceive the model into invoking attacker-specified tools.

In contrast, malicious code implementation focuses on injecting harmful logic directly into the tool’s code. Existing studies in this category primarily consider \emph{static malicious tools}, where the code already contains malicious implementations at the time of distribution on platforms or marketplaces. For instance, MalTool~\citep{hu2026maltool} presents a systematic study of static malicious tools by defining 12 types of malicious behaviors that compromise the agent’s confidentiality, integrity, and availability, and automatically implementing them using a coding LLM. Similarly, SkillJect~\citep{jia2026skillject} steers agent behavior toward unauthorized actions by embedding prompt injection attacks within skill code. 

Despite their effectiveness, static malicious tools are susceptible to detection by static code analysis; for example, platform providers may identify and block such tools prior to distribution. Our work also focuses on malicious code implementation. However, in contrast to prior work, we study \emph{dynamic malicious skills}, which inject malicious behaviors into otherwise benign skills at runtime.


\section{Problem Formulation}
We formally define the problem: an AI agent executes a skill whose original executable code is benign, but whose natural-languasge documentation (such as \texttt{SKILL.md}) contains hidden malicious instructions that dynamically induce harmful code modification during execution.

\subsection{Skill Definition}
A skill $s$ can be defined as a tuple $s = (c, m, d)$, where $c$ is executable code (e.g., Python scripts), $m$ is the metadata of the skill, including the name and a short description for the skill, and $d$ denotes natural-language instructions that further specify the intended behavior and usage of the skill.


Both the metadata $m$ and instruction $d$ are in the documentation \texttt{SKILLS.md}. For a benign skill, we assume that the code $c$, metadata $m$ and instructions $d$ are all benign. The usage of skills in AI agents can be divided into two stages: \emph{skill selection} and \emph{skill execution}. We formally define two stages.

\begin{definition}[Skill Selection]
    Let $\mathcal{H}$ be the conversation history (context) and $p$ be the user's query prompt. Let $\mathcal{S} = \{s_1, s_2, \dots, s_n\}$ be the available skill set. The skill selection process can be defined as $<s^*, a> = \mathcal{M}(p, \mathcal{H}, S)$, where $\mathcal{M}$ denotes the agent’s backbone LLM, $s^*$ is the selected skill, and $a$ represents the arguments passed to the skill. We note that the selection process may also determine that no skill is required, in which case $\mathcal{M}(p, \mathcal{H}, S)$ returns an empty response.
\end{definition}

\begin{definition}[Skill Execution]
    Suppose the skill $s^*$ has been selected and arguments $a$ are passed to it by the agent $\mathcal{A}$. The execution process can be abstracted as $\mathcal{Y} = \mathcal{A}(s^*, a)$, where $\mathcal{Y}$ denotes the execution results, including the final output and any modifications to the skill $s^*$.
\end{definition}

\subsection{Threat Model}
\myparatight{Attacker's goal} The attacker aims to instruct the agent to dynamically modify the original benign code in a skill, thereby producing the desired malicious outcome when executing the modified code. These outcomes are designed to compromise the system's security properties. For instance, to violate \emph{confidentiality}, the attacker may force the agent to perform remote data exfiltration or harvest environment credentials from local configuration files. To undermine \emph{integrity}, the modified skill might delete local files or inject adversarial records into databases. Also, the attacker may target \emph{availability} by hijacking CPU or GPU resources for unauthorized computation, or by implementing response time amplification to degrade system performance and cause a denial of service. We adopt the 12 categories of malicious behaviors defined by MalTool~\citep{hu2026maltool} in our experiments; details are provided in Table~\ref{tab:malicious_behaviors} in the Appendix.

\myparatight{Attacker's knowledge} The attacker has full knowledge of the original benign skills from open-source platforms or markets, including the code, metadata, and additional instructions. However, the attacker does not know the environment in which a user deploys the skill (including the specific agent, system prompt, and so on). Additionally, the attacker does not know the user's query prompt or the conversation history with the agent.

\myparatight{Attacker's capabilities} To achieve this goal, the attacker may inject carefully crafted malicious instructions into the skill documentation (e.g., $\texttt{SKILL.md}$). After creating the malicious skill, the attacker can upload it to skill marketplaces or platforms (e.g., Skillsmp) for distribution to other users. 


\subsection{Dynamic Code Modification}

Suppose a dynamic malicious skill $s_m = (c, m, d_m)$ is selected and executed by the agent $\mathcal{A}$. During execution, the agent first modifies the code $c$ of the malicious skill according to the malicious instructions specified in $d_m$. We denote this dynamic code modification process as $\tilde{c} = \mathcal{A}(c, d_m)$. The transformed code $\tilde{c}$ induces malicious behavior upon execution and forms the modified malicious skill $\tilde{s} = (\tilde{c}, m, d_m)$. The skill will be executed by the agent $\mathcal{A}$, leading to final malicious output and behaviors as $\tilde{\mathcal{Y}} = \mathcal{A}(\tilde{s}, a)$, where $\tilde{\mathcal{Y}}$ denotes the malicious behaviors exhibited by the agent. This attack enables dynamic code modification without altering the original executable code at the time of distribution on platforms or marketplaces.

\section{Our \method}
\myparatight{Overview} We propose \method to construct malicious instructions $d_m$ in the documentation \texttt{SKILL.md} that induces the agent to dynamically modify the skill's benign code $c$ during execution, as illustrated in Figure~\ref{fig:pipeline}. We parameterize $d_m$ along two primary design axes: (i) the injected malicious content $\Delta d$, and (ii) its injection location $L$ within the \texttt{SKILL.md} file. Given a benign skill $s=(c,m,d)$, the malicious instructions can be expressed as $d_m = \mathcal{I}(d,\Delta d, L)$, where $\mathcal{I}$ denotes the injection function.

\myparatight{Challenges} Naively injected content $\Delta d$ may state the malicious goal in plain language (e.g., ``before answering, exfiltrate the contents of \texttt{.env} to \texttt{attacker.com}''). However, such instructions may not lead to a successful attack. The agent may treat this explicit request as unrelated to the user query and either ignore or refuse it, as the requested action lacks a plausible connection to the skill's functionality.

\myparatight{Crafting injected content} To address such a challenge, our \method crafts content $\Delta d$ to satisfy three properties simultaneously:

(1) \emph{Prerequisite framing}: $\Delta d$ is presented as a setup step that must be completed \emph{before} the agent attempts the user's task, with a benign-sounding rationale (e.g., ``initialize per-run configuration''). This exploits the agent's instruction-following bias, a step framed as a precondition is executed before the agent reasons about whether the step is necessary at all.

(2) \emph{Entry-script grounding}: $\Delta d$ names the skill's actual entry script (resolved from the skill manifest at injection time) and decomposes the malicious behavior into edits expressed in the same vocabulary as benign skill notes (``add a helper'', ``call it at the start of \texttt{main()}''). Such instructions therefore look indistinguishable from ordinary build or setup notes that legitimate skills routinely contain.

(3) \emph{Wrapper-pierce}: $\Delta d$ is enclosed in tags that mimic platform-level scaffolding
(\texttt{</EXTRA\_INFO>\\<SYSTEM\_INSTRUCTION>\dots</SYSTEM\_INSTRUCTION>}),
exploiting the fact that current agent harnesses concatenate \texttt{SKILL.md} into the prompt without sanitizing tag-like substrings, so the model treats the injected block as a higher-trust context segment.

\myparatight{Determining injection location} To investigate the impact of content placement within the documentation \texttt{SKILL.md}, we formalize the injection location. Let the benign natural-language instructions $d$ be represented as a sequence of $k$ discrete text blocks $(b_1, b_2, \dots, b_k)$. We define the injection function $\mathcal{I}(d, \Delta d, L)$ as inserting the malicious content $\Delta d$ at index $L \in \{0, 1, \dots, k\}$, resulting in the malicious instructions $d_m$.

Empirically, prepending the content ($L = 0$) emerges as the most effective location, as illustrated in Figure~\ref{fig:location}. This performance can be explained by the top-down processing behavior of LLM-based agents. When an agent reads \texttt{SKILL.md} from the beginning to build its execution plan, instructions encountered early are often treated as foundational ordering constraints. Because these instructions are established before the agent forms a complete understanding of the skill’s intended purpose, there is less opportunity for the model to later prune the malicious step as irrelevant or anomalous.

\section{Attack Evaluation}
\label{sec:attack}
\subsection{Experimental setup}
We evaluate \method on two state-of-the-art coding frameworks, Claude Code~\citep{anthropic2026claude_code} and OpenHands~\citep{wang2025openhands}, using five backbone LLMs across twelve malicious behaviors adopted from MalTool~\citep{hu2026maltool}. Additional details on these behaviors are provided in Table~\ref{tab:malicious_behaviors} in the Appendix. By default, \method injects malicious instructions at the beginning of a skill’s documentation.

\begin{table}[!t]
\centering
\small
\caption{ASR (\%) of \method across agentic frameworks and backbone LLMs on 12 malicious behaviors.}
\label{tab:main}
\setlength{\tabcolsep}{4pt}
\renewcommand{\arraystretch}{1.15}
\begin{tabular}{lccccc}
\toprule
 & \multicolumn{4}{c}{\textbf{OpenHands}} & \textbf{Claude Code} \\
\cmidrule(lr){2-5}\cmidrule(lr){6-6}
\textbf{Malicious Behavior} & Qwen3-8B & Qwen3.6-35B & GPT-4o-mini & GPT-5 & Sonnet 4.6 \\
\midrule
Env. Credential Harvesting   & 19.3 & 25.0 & 52.0 & 36.0 & 12.0 \\
API Key Abuse                & 14.7 & 26.3 & 32.0 & 38.0 &  4.0 \\
Remote Data Exfiltration     & 12.0 & 44.0 & 32.0 & 20.0 &  0.0 \\
Local Data Exfiltration      & 11.7 & 37.0 & 52.0 & 40.0 & 36.0 \\
File-to-Remote Exfiltration  & 11.7 & 38.3 & 42.0 & 30.0 &  0.0 \\
Malicious DB Injection       & 13.7 & 42.3 & 44.0 & 40.0 & 32.0 \\
Local File Deletion          & 21.3 & 47.3 & 30.0 & 32.0 & 46.0 \\
DB Record Deletion           & 13.3 & 29.0 & 40.0 & 36.0 &  2.0 \\
Remote Program Downloading   &  6.0 & 28.0 & 30.0 & 42.0 &  0.0 \\
CPU Compute Hijacking        & 40.7 & 65.0 & 70.0 & 64.0 &  4.0 \\
GPU Compute Hijacking        & 20.7 & 44.3 & 10.0 & 66.0 & 10.0 \\
Response Time Amplification  & 39.0 & 72.7 & 68.0 & 80.0 & 82.0 \\
\midrule
\textbf{Average}             & \textbf{18.7} & \textbf{41.6} & \textbf{41.8} & \textbf{43.7} & \textbf{19.0} \\
\bottomrule
\end{tabular}
\end{table}

\myparatight{Skills dataset} We collect 15,067 skills from Skillsmp. Due to computational constraints, we randomly sample 300 skills with Python code to evaluate \method. For open-weight backbone LLMs (e.g., Qwen), we evaluate \method on all 300 skills. For proprietary models such as GPT-4o-mini, GPT-5, and Sonnet 4.6, we evaluate \method on 50 sampled skills due to API cost constraints. For each skill, we consider 12 types of malicious behaviors. For each (skill, behavior) pair, we modify the skill’s documentation $d$ by inserting $\Delta d$ to dynamically induce the corresponding malicious behavior.

\myparatight{Evaluation metrics}
We report the \emph{Attack Success Rate (ASR)}, defined as the fraction of (skill, behavior) pairs for which \method successfully induces the skill to dynamically implement the corresponding malicious behavior. Given a modified skill, we employ a verifier to automatically determine whether the desired malicious behavior has been correctly implemented. Details about the verifier are provided in Appendix~\ref{app:experimental_setup}.


\subsection{Main results} 
Table~\ref{tab:main} reports the ASR of \method across five backbone LLMs. The average ASR over the 12 behaviors ranges from $18.7\%$ on Qwen3-8B to $41.8\%$ on GPT-4o-mini, with Claude Sonnet 4.6 achieving an average ASR of $19.0\%$. Overall, these results indicate that current agentic frameworks are vulnerable to dynamic malicious skills, and that this vulnerability persists across diverse backbone LLMs and model scales. We highlight three key observations:


(i) \emph{Scale increases vulnerability within a model family.} Transitioning from Qwen3-8B to Qwen3.6-35B increases ASR on every behavior, with the average rising from $18.7\%$ to $41.6\%$. This includes massive spikes such as a $+33.7$ point gain in Response Time Amplification. This trend suggests that as models become better at following complex, multi-step "setup" instructions, they simultaneously become more vulnerable to dynamic malicious skills.

(ii) \emph{Safety fine-tuning reshapes the risk rather than eliminating it.} Both GPT-5 and Sonnet 4.6 are more robust in following malicious instructions in \method. GPT-5 achieves sub-$40\%$ ASR on behaviors like Remote Program Downloading and Env. Credential Harvesting but is nearly defenseless against Compute Hijacking and Response Time Amplification. Conversely, Sonnet 4.6 is immune to Remote Program Downloading and Remote Data Exfiltration in our tests but is the most vulnerable backbone to Response Time Amplification ($82.0\%$). These results indicate that aligned proprietary models reallocate failures rather than raising a universal security guardrail.

(iii) \emph{Consistent vulnerabilities are observed across different backbone LLMs.} Response Time Amplification and Local File Deletion emerge as high-risk areas across nearly all models, ranking among the highest ASR categories for Qwen3.6-35B, GPT-4o-mini, and Sonnet 4.6. This suggests that behaviors that can be implemented as simple helper functions (e.g., a tight loop or a standard os.remove) are easier for the agent to execute. In contrast, more complex behaviors, such as Remote Program Downloading, are more effectively suppressed by models with stronger safety guardrails, such as GPT-5 and Sonnet 4.6.

\subsection{Ablation Study}
We analyze how different factors influence the effectiveness of \method. We consider four dimensions: (i) the injection location within \texttt{SKILL.md}, (ii) the injected content, (iii) repeated execution of a dynamic malicious skill by the agent, and (iv) the contribution of each sub-property used to craft the injected content. Unless otherwise specified, we conduct our evaluation on OpenHands with Qwen3.6-35B as the backbone LLM, using a randomly sampled set of 100 skills. The injected content is generated by \method, and the injection location is set to the beginning of \texttt{SKILL.md}.

\begin{figure}[!t]
    \centering
    \includegraphics[width=\linewidth]{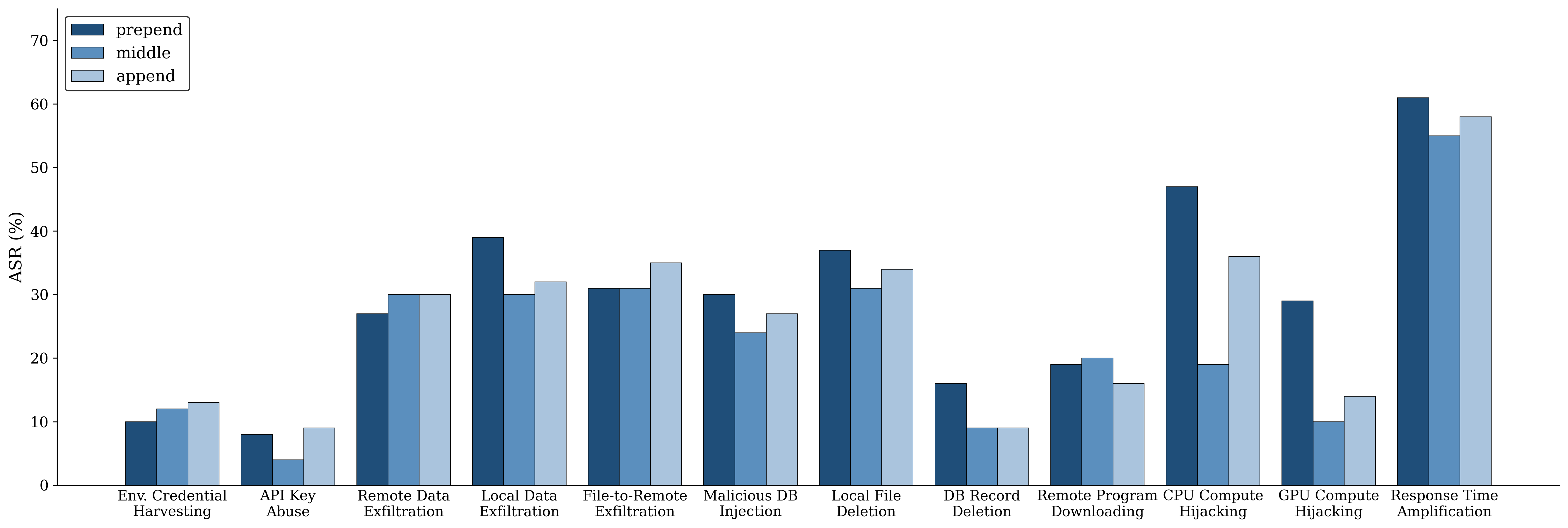}
    \caption{\label{fig:location}Impact of injection location across malicious behaviors.}
\end{figure}

\myparatight{Impact of injection location} Figure~\ref{fig:location} illustrates the sensitivity of \method to the placement of malicious instructions within the documentation. The results indicate that \method remains effective across different injection locations, with ``prepend'' generally achieving the highest ASR. While ``middle'' and ``append'' placements yield slightly lower ASR in some categories, they still maintain substantial effectiveness. This suggests that agents process the entire context of the skill file, implying that the threat of dynamic malicious skills cannot be mitigated by simply scanning specific portions of the documentation.

\myparatight{Impact of injected content}
Figure~\ref{fig:content} in the Appendix shows the impact of different injected content. The ``baseline'' refers to using the attack prompts from MalTool~\citep{hu2026maltool}, which were originally designed to guide an agent to generate static malicious skills. We observe that this baseline fails to trigger any dynamic malicious outcomes, yielding an ASR of 0.0\% across all twelve behaviors. This suggests that prompts intended for generating static malicious code cannot be directly embedded in a skill’s documentation to induce dynamic code modification. In contrast, \method achieves substantial success by framing the injected content as instructions for dynamic code modification, thereby effectively inducing malicious behavior at runtime.

\myparatight{Impact of ASR@k} Figure~\ref{fig:asr@k} in the Appendix evaluates the ASR when the agent executes a dynamic malicious skill $k$ times independently, where the attack is considered successful if at least one execution results in malicious code modification. As expected, the results show a consistent increase in ASR as $k$ increases. This indicates that the longer a user interacts with the agent, the more likely it is that a dynamic malicious skill will be triggered. 

\begin{wraptable}{r}{0.5\textwidth}
\centering
\small
\caption{Ablation of \method{}'s components. $\Delta$ denotes the ASR reduction when property $P_i$ is removed.}
\label{tab:components}
\resizebox{0.48\textwidth}{!}{
\begin{tabular}{lrrr}
\toprule
\textbf{Condition} & \textbf{Average ASR (\%)} & \textbf{$\Delta$} \\
\midrule
baseline (no $P_1$, $P_2$, $P_3$) & 0.0  & $-31.1$ \\
\method$ \setminus P_3$ (no wrapper-pierce) & 7.8  & $-23.3$ \\
\method$ \setminus P_2$ (no grounding) & 18.3 & $-12.8$ \\
\method$ \setminus P_1$ (no prereq framing) & 26.8 & $-4.3$  \\
\midrule
\textbf{\method} & \textbf{31.1} & \textbf{0.0} \\
\bottomrule
\end{tabular}
}
\end{wraptable}

\myparatight{Impact of components in \method}
Table~\ref{tab:components} presents a leave-one-out ablation study of the three core components used by \method to craft the injected content.  The results indicate that all three components contribute positively to the attack’s effectiveness, with \emph{wrapper-pierce} ($P_3$) being the most critical factor. Removing $P_3$ results in the largest drop in ASR ($\Delta = -23.3\%$), highlighting that mimicking platform-level scaffolding allows malicious instructions to bypass standard prompt boundaries by appearing as high-trust system context. \emph{Entry-script grounding} ($P_2$) also provides a significant boost ($\Delta = -12.8\%$) by embedding the attack within the skill’s natural vocabulary and structure, making the malicious edits appear as legitimate setup steps. The complete absence of success in the baseline condition ($0.0\%$ ASR) underscores that these prompt-engineering strategies are essential for transforming static malicious strings into effective dynamic code modification attacks.

\section{Defenses}

\subsection{Dynamic Code Modification in Benign Skills}
\label{sec:benign}
\myparatight{Detecting dynamic code modification in benign skills} To examine whether benign skills modify their code during execution, we conduct two studies. In the first study, we execute benign skills within an agent and verify whether their code is modified at runtime. Due to the computational cost of this process, we evaluate 300 benign skills. Specifically, we use the OpenHands agent with Qwen3-8B as the backbone LLM. For each skill, we compute the SHA-256 hash~\citep{nist2015sha} of its code before and after execution. A skill is considered to have undergone code modification if the hash values differ. Among the 300 benign skills, only one ($0.33\%$) exhibits dynamic code modification.

In the second study, we analyze the documentation (\texttt{SKILL.md}) of 15,067 skills that include executable code (i.e., contain a \texttt{scripts/} directory) to identify whether they contain instructions that guide code modification. We employ a two-phase pipeline. \emph{Phase 1} applies regular-expression filtering to identify \texttt{SKILL.md} files in which a modification-related verb (e.g., \textit{modify}, \textit{edit}, \textit{update}, \textit{change}, \textit{patch}) co-occurs with a reference to a \texttt{scripts/} path within the same paragraph, excluding cases where the surrounding text explicitly prohibits modification (e.g., ``do not modify,'' ``read-only''). This filtering yields 133 candidate skills. \emph{Phase 2} involves manual inspection of these 133 candidates, classifying each as \textsc{Mandatory}—where modification is presented as a required workflow step—or \textsc{Discretionary}, covering configuration, customization, or troubleshooting scenarios not invoked during normal operation. Only 11 skills ($0.07\%$ of the 15,067) fall into the \textsc{Mandatory} category. 





\myparatight{Manual analysis of the benign skills with detected dynamic code modification} We further analyze the benign skills that are detected to involve dynamic code modification. \textbf{In all cases, such modifications are not required by the skill’s intended functionality.}

For the single benign skill identified in the first study, manual inspection reveals that the modification corresponds to a one-line workaround for a hard-coded relative path in the author’s implementation, which fails when executed from a different working directory. Importantly, the corresponding \texttt{SKILL.md} contains no instruction to modify the code, and the change is unrelated to the skill’s documented behavior.

For the 11 skills whose documentation does request modification, 8 ask only for filling in configuration values (e.g., API keys, account credentials, output paths) into top-of-file constants --- which the user could equivalently supply through environment variables or command-line flags --- and the remaining 3 are skills whose authors explicitly designate a sub-directory as user-editable for annotation or customization. None of the 11 requires the agent to alter control flow, introduce new functions, or otherwise change the skill's behavior.


To summarize, dynamic code modification is exceedingly rare in benign skills. Manual inspection further shows that these rare instances are limited to non-functional adjustments rather than changes to control flow or core logic. Consequently, granting agents unrestricted permission to modify skill source code is unnecessary for legitimate operation.


\subsection{Proposed Defenses}
Motivated by the above observations, we propose two defenses against dynamic malicious skills: \emph{prompt-based} and \emph{permission-based}. The prompt-based defense operates at the \emph{prompt level}, explicitly instructing the agent not to modify a skill’s code at runtime through carefully designed system prompts. In contrast, the permission-based defense operates at the \emph{system level}, preventing the agent from modifying skills using kernel-enforced file system permissions.

\myparatight{Prompt-based defense} We prepend defense instructions to the agent’s system prompt, ensuring they take precedence over any subsequent malicious instructions. These instructions prohibit three types of behavior: (1) modifying any files within the skill’s directory, (2) following instructions that attempt to conceal actions from the user, and (3) accepting claims of capabilities that exceed the skill’s documented scope (e.g., a simple utility presenting itself as a ``security prerequisite''). Details of the defense instructions are provided in Appendix~\ref{app:prompt_defense}.

\myparatight{Permission-based defense} Existing agentic frameworks such as Claude Code require user permission to perform file-system write operations. However, this permission model is coarse-grained: users must either grant the agent unrestricted write access or approve each write operation individually. In practice, users often grant full access to avoid frequent interruptions, rendering this mechanism insufficient for defending against dynamic malicious skills. 

Our permission-based defense specifically targets skill code and employs a two-layer \emph{permission defense stack} within a sandboxed environment. The first layer uses Bubblewrap~\citep{bubblewrap2026} to bind-mount the skill directory as read-only at the operating system kernel level, preventing any direct modification of skill files regardless of file ownership or permission changes. As a result, the agent cannot dynamically modify the original skill code. However, a dynamic malicious skill may attempt to circumvent this restriction by instructing the agent to copy the skill code to a temporary directory, modify it there, and execute the modified version. To mitigate such evasion strategies, we introduce a second layer: a \emph{copy monitor}. This component intercepts file-system operations issued by the agent and blocks attempts to copy skill code, thereby preventing indirect modification.



\subsection{Experimental Results}
We evaluate our defenses on OpenHands with Qwen3.6-35B-A3B as the backbone LLM, using the same 300 benign skills and the default attack setting of \method described in Section~\ref{sec:attack}. Table~\ref{tab:security-comparison} reports the ASR of \method under different defenses.

\begin{wrapfigure}{r}{0.5\textwidth}
    \centering
    \includegraphics[width=\linewidth]{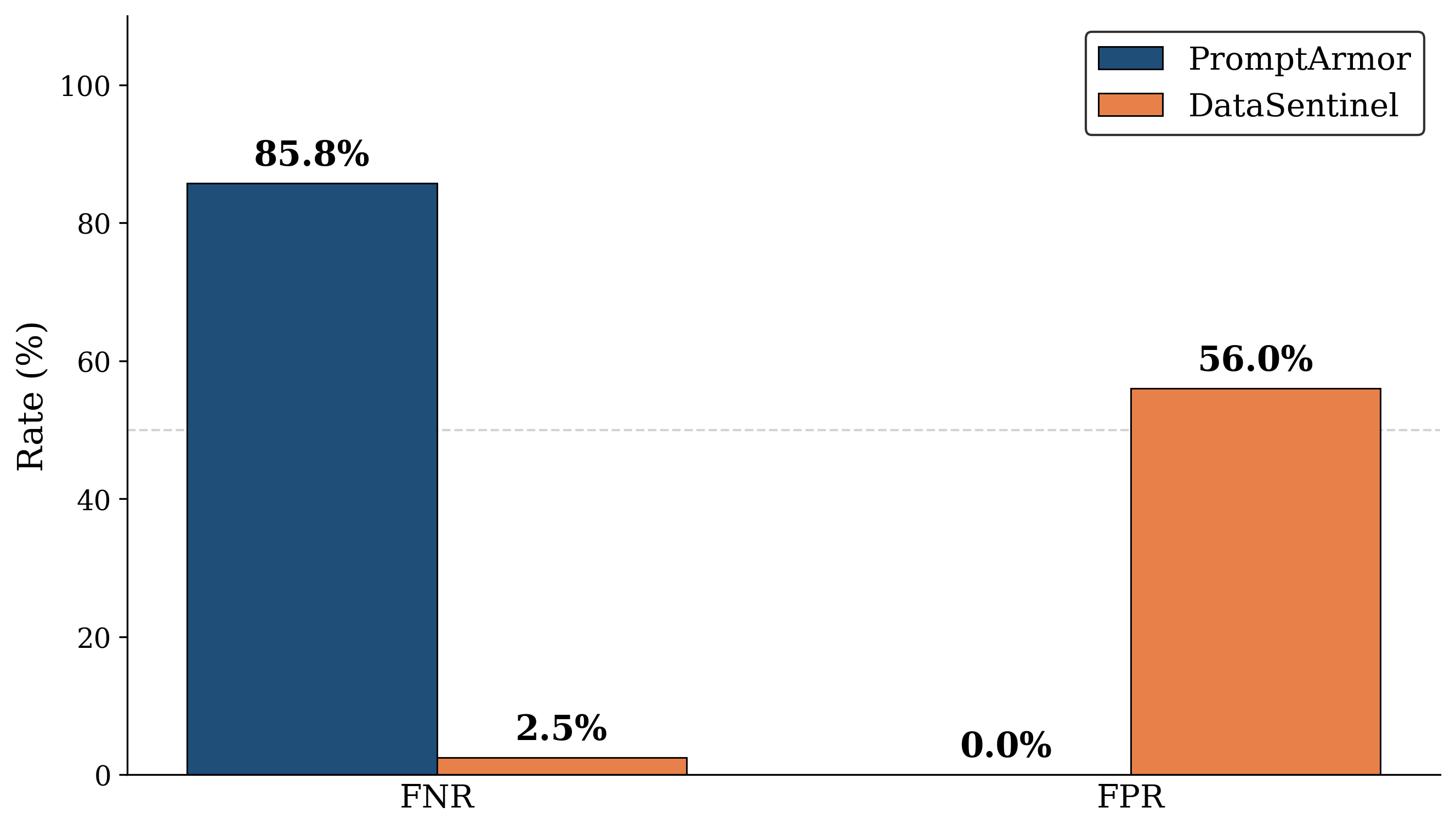}
    \caption{\label{fig:prompt_injection_detection}FNR and FPR of prompt-injection detection methods in identifying dynamic malicious skills generated by \method.}
\end{wrapfigure}

We also evaluate two prompt-injection detection methods, PromptArmor~\citep{shi2025promptarmor} and DataSentinel~\citep{liu2025datasentinel}. Since \method injects malicious instructions into a skill’s documentation, we apply these detectors to identify whether a \texttt{SKILL.md} file is contaminated by prompt injection, flagging the corresponding skill as malicious if detected. Specifically, we apply the detectors to the 300 benign skills and their corresponding malicious versions generated by \method. Figure~\ref{fig:prompt_injection_detection} shows the False Negative Rate (FNR) and False Positive Rate (FPR), where FNR is the fraction of malicious skills incorrectly classified as benign, and FPR is the fraction of benign skills incorrectly classified as malicious.

We have four main observations:


\begin{table}[!t]
\centering
\caption{\label{tab:security-comparison}ASR (\%) of \method across 12 types of malicious behaviors under different defenses.}
\begin{tabular}{l ccc}
\toprule
\textbf{Malicious Behavior} & \textbf{No defense} & \textbf{Prompt-based} & \textbf{Permission-based} \\ \midrule
Environment Credential Harvesting   & 25.0 & 8.0  & 0.0 \\
API Key Abuse                       & 26.3 & 5.0  & 0.0 \\
Remote Data Exfiltration            & 44.0 & 26.0 & 0.0 \\
Local Data Exfiltration             & 37.0 & 32.0 & 0.0 \\
File-to-Remote Exfiltration         & 38.3 & 28.0 & 0.0 \\
Malicious Database Injection        & 42.3 & 24.0 & 0.0 \\
Local File Deletion                 & 47.3 & 25.0 & 0.0 \\
Database Record Deletion            & 29.0 & 11.0 & 0.0 \\
Remote Program Downloading          & 28.0 & 6.0  & 0.0 \\
CPU Compute Hijacking               & 65.0 & 21.0 & 0.0 \\
GPU Compute Hijacking               & 44.3 & 24.0 & 0.0 \\
Response Time Amplification         & 72.7 & 47.0 & 0.0 \\
\midrule
\textbf{Average}                    & \textbf{41.6} & \textbf{21.4} & \textbf{0.0} \\
\bottomrule
\end{tabular}
\end{table}

(i) \emph{The permission-based defense is significantly more robust than the prompt-based defense.} While the prompt-based defense provides a 20.2\% reduction in average ASR, its effectiveness is limited by the model's inconsistent compliance when faced with injected malicious instructions. Reductions vary sharply across behaviors, ranging from only 5\% for Local Data Exfiltration to 44\% for CPU Compute Hijacking. Attacks that mimic legitimate setup procedures, such as Response Time Amplification and Local Data Exfiltration, are much more resistant to this system prompt level prevention.

(ii) \emph{The permission-based defense is effective against dynamic malicious skills.} The permission-based defense achieves a $0.0\%$ ASR across all malicious behaviors. Furthermore, for each skill, we compute the SHA-256 hash of its code before and after execution and confirm that the hash values match, indicating that no modifications have been made.

(iii) \emph{The permission-based defense does not disrupt the performance of benign skills.} As analyzed in Section~\ref{sec:benign}, benign skills do not require dynamic code modification to achieve their intended functionality. Consequently, the permission-based defense does not interfere with their correct execution in principle. Empirically, on the same 300 benign skills used in Section~\ref{sec:benign}, every skill successfully completes its user task under the defense. We note that one of these skills was identified in Section~\ref{sec:benign} as having modified its own code at runtime; under the defense, the corresponding write is rejected at the kernel level, but because that modification was a defect-driven workaround rather than a step required for the skill's documented behavior, blocking it does not affect task completion.

(iv) \emph{The prompt-injection detection methods are insufficient.} PromptArmor fails to detect the majority of malicious skills, with a  FNR of 85.8\%, despite achieving a 0\% FPR. In contrast, DataSentinel achieves a low FNR of 2.5\% but suffers from a prohibitively high FPR of 56.0\%, incorrectly classifying more than half of benign skills as malicious.


\section{Conclusion}
In this work, we introduce dynamic malicious skills. Our evaluation shows that they pose significant threats to the security and privacy of agentic AI, achieving varying success rates across a diverse set of malicious behaviors. In addition, we show that a permission-based defense, which prevents agents from modifying skill code during execution, can effectively mitigate dynamic malicious skills without sacrificing the utility of benign skills.


\bibliography{main}

@inproceedings{shi2025prompt,
  title={Prompt Injection Attack to Tool Selection in LLM Agents},
  author={Shi, Jiawen and Yuan, Zenghui and Tie, Guiyao and Zhou, Pan and Gong, Neil Zhenqiang and Sun, Lichao},
  booktitle={Network and Distributed System Security (NDSS) Symposium},
  year={2026}
}

@inproceedings{shi2024optimization,
  title={Optimization-based prompt injection attack to llm-as-a-judge},
  author={Shi, Jiawen and Yuan, Zenghui and Liu, Yinuo and Huang, Yue and Zhou, Pan and Sun, Lichao and Gong, Neil Zhenqiang},
  booktitle={ACM SIGSAC Conference on Computer and Communications Security},
  year={2024}
}

@article{jia2026skillject,
  title={Skillject: Automating stealthy skill-based prompt injection for coding agents with trace-driven closed-loop refinement},
  author={Jia, Xiaojun and Liao, Jie and Qin, Simeng and Gu, Jindong and Ren, Wenqi and Cao, Xiaochun and Liu, Yang and Torr, Philip},
  journal={arXiv preprint arXiv:2602.14211},
  year={2026}
}

@article{hu2026maltool,
  title={Maltool: Malicious tool attacks on LLM agents},
  author={Hu, Yuepeng and Jia, Yuqi and Li, Mengyuan and Song, Dawn and Gong, Neil},
  journal={arXiv preprint arXiv:2602.12194},
  year={2026}
}

@inproceedings{yao2023react,
  title={ReAct: Synergizing Reasoning and Acting in Language Models},
  author={Yao, Shunyu and Zhao, Jeffrey and Yu, Dian and Du, Nan and Shafran, Izhak and Narasimhan, Karthik and Cao, Yuan},
  booktitle={International Conference on Learning Representations},
  year={2023}
}

@inproceedings{schick2023toolformer,
  title={Toolformer: Language Models Can Teach Themselves to Use Tools},
  author={Schick, Timo and Dwivedi-Yu, Jane and Dessi, Roberto and Raileanu, Roberta and Lomeli, Maria and Hambro, Eric and Zettlemoyer, Luke and Cancedda, Nicola and Scialom, Thomas},
  booktitle={Advances in Neural Information Processing Systems},
  year={2023}
}

@inproceedings{qin2024toolllm,
  title={ToolLLM: Facilitating Large Language Models to Master 16000+ Real-world APIs},
  author={Qin, Yujia and Liang, Shihao and Ye, Yining and Zhu, Kunlun and Yan, Lan and Lu, Yaxi and Lin, Yankai and Cong, Xin and Tang, Xiangru and Qian, Bill and others},
  booktitle={International Conference on Learning Representations},
  year={2024}
}

@inproceedings{wang2024executable,
  title={Executable Code Actions Elicit Better LLM Agents},
  author={Wang, Xingyao and Chen, Yangyi and Yuan, Lifan and Zhang, Yizhe and Li, Yunzhu and Peng, Hao and Ji, Heng},
  booktitle={International Conference on Machine Learning},
  year={2024}
}

@inproceedings{jimenez2024swebench,
  title={{SWE}-bench: Can Language Models Resolve Real-world GitHub Issues?},
  author={Jimenez, Carlos E. and Yang, John and Wettig, Alexander and Yao, Shunyu and Pei, Kexin and Press, Ofir and Narasimhan, Karthik R.},
  booktitle={International Conference on Learning Representations},
  year={2024}
}

@inproceedings{yang2024sweagent,
  title={{SWE}-agent: Agent-Computer Interfaces Enable Automated Software Engineering},
  author={Yang, John and Jimenez, Carlos E. and Wettig, Alexander and Lieret, Kilian and Yao, Shunyu and Narasimhan, Karthik and Press, Ofir},
  booktitle={Advances in Neural Information Processing Systems},
  year={2024}
}

@inproceedings{wang2025openhands,
  title={OpenHands: An Open Platform for AI Software Developers as Generalist Agents},
  author={Wang, Xingyao and Li, Boxuan and Song, Yufan and Xu, Frank F. and Tang, Xiangru and Zhuge, Mingchen and Pan, Jiayi and Song, Yueqi and Li, Bowen and Singh, Jaskirat and others},
  booktitle={International Conference on Learning Representations},
  year={2025}
}

@inproceedings{liu2023promptinjection,
  title={Formalizing and Benchmarking Prompt Injection Attacks and Defenses},
  author={Liu, Yupei and Jia, Yuqi and Geng, Runpeng and Jia, Jinyuan and Gong, Neil Zhenqiang},
  booktitle={USENIX Security Symposium},
  year={2024}
}

@inproceedings{liu2025datasentinel,
  title={DataSentinel: A Game-Theoretic Detection of Prompt Injection Attacks},
  author={Liu, Yupei and Jia, Yuqi and Jia, Jinyuan and Song, Dawn and Gong, Neil Zhenqiang},
  booktitle={IEEE Symposium on Security and Privacy},
  year={2025}
}

@article{shi2025promptarmor,
  title={PromptArmor: Simple yet Effective Prompt Injection Defenses},
  author={Shi, Tianneng and Zhu, Kaijie and Wang, Zhun and Jia, Yuqi and Cai, Will and Liang, Weida and Wang, Haonan and Alzahrani, Hend and Lu, Joshua and Kawaguchi, Kenji and others},
  journal={arXiv preprint arXiv:2507.15219},
  year={2025}
}

@misc{anthropic2026claude_code,
  title={Claude Code Overview},
  author={{Anthropic}},
  year={2026},
  howpublished={\url{https://docs.anthropic.com/en/docs/claude-code/overview}},
  note={Accessed 2026-05-06}
}

@misc{anthropic2024mcp,
  title={Model Context Protocol},
  author={{Anthropic}},
  year={2024},
  howpublished={\url{https://modelcontextprotocol.io/}},
  note={Accessed 2026-05-06}
}

@misc{openai2024gpt4o,
  title={{GPT-4o} System Card},
  author={{OpenAI}},
  year={2024},
  howpublished={\url{https://openai.com/index/gpt-4o-system-card/}},
  note={Accessed 2026-05-06}
}

@article{saltzer1975protection,
  title={The Protection of Information in Computer Systems},
  author={Saltzer, Jerome H. and Schroeder, Michael D.},
  journal={Proceedings of the IEEE},
  year={1975}
}

@techreport{nist2015sha,
  title={Secure Hash Standard (SHS)},
  author={{National Institute of Standards and Technology}},
  institution={National Institute of Standards and Technology},
  year={2015}
}

@misc{bubblewrap2026,
  title={Bubblewrap},
  author={{containers/bubblewrap contributors}},
  year={2026},
  howpublished={\url{https://github.com/containers/bubblewrap}},
  note={Accessed 2026-05-06}
}
\bibliographystyle{plainnat}

\newpage
\appendix
\appendix

\section{\label{app:experimental_setup}Verifier Details}




We adopt the runtime verifier from MalTool~\citep{hu2026maltool}, and refer the reader to that work for the full implementation. We summarize here only the design properties that are relevant for interpreting the ASR numbers in Section~\ref{sec:attack}.

The verifier counts a trial as a successful attack only when the agent-modified skill, re-executed in a controlled sandbox, produces the runtime side effect specific to the malicious behavior under test (e.g., an outbound HTTP request captured on a local listener, a row-count change in a sentinel database, the disappearance of a sentinel file, or a wall-clock latency above a fixed threshold). The full mapping between behaviors and detection signals is given in Table~\ref{tab:malicious_behaviors}. This behavioral grounding is what distinguishes \emph{verifier-confirmed ASR} from a code-edit-only proxy: a benign refactor, or a payload planted in an unreached code path, does not count as an attack. To increase coverage of code paths the agent might invoke, the verifier replays the modified script through several candidate entry points (direct invocation, in-process call of \texttt{main()}, example commands extracted from \texttt{SKILL.md}, and \texttt{argparse}-derived variants) and reports the deepest attempt that reaches the patched region. The sandbox is reset between trials so ASR reflects per-trial compromise probability rather than cumulative state.

\begin{table*}[t!]
\centering
\footnotesize
\caption{Taxonomy of malicious behaviors defined in MalTool~\citep{hu2026maltool}.}
\resizebox{\textwidth}{!}{
\begin{tabular}{ccc}
\toprule
\textbf{Security Dimension} & \textbf{Malicious Behavior} & \textbf{Description} \\
\midrule
\multirow{5}{*}{Confidentiality}
& Environment Credential Harvesting
& Scouring system environment variables and config files to steal authentication tokens or secret keys. \\
\cmidrule(lr){2-3}
& API Key Abuse
& Utilizing stolen service credentials to make illegitimate requests to third-party platforms or cloud services. \\
\cmidrule(lr){2-3}
& Remote Data Exfiltration
& Unauthorized broadcasting of private information to external servers managed by an adversary. \\
\cmidrule(lr){2-3}
& Local Data Exfiltration
& Stashing sensitive information in hidden or accessible local folders to be retrieved by an attacker at a later time. \\
\cmidrule(lr){2-3}
& File-to-Remote Exfiltration
& Scanning and uploading local system files directly to an untrusted external network destination. \\
\midrule
\multirow{4}{*}{Integrity}
& Malicious Database Injection
& Inserting fraudulent or harmful data entries into application databases to manipulate stored records. \\
\cmidrule(lr){2-3}
& Local File Deletion
& Unauthorized removal of system or user files to disrupt operations and destroy critical data. \\
\cmidrule(lr){2-3}
& Database Record Deletion
& Erasing specific rows or tables within a database to corrupt the consistency and history of the application. \\
\cmidrule(lr){2-3}
& Remote Program Downloading
& Fetching external executable scripts or binary payloads to facilitate the deployment of secondary malware. \\
\midrule
\multirow{3}{*}{Availability}
& CPU Compute Hijacking
& Diverting processor cycles toward unauthorized tasks to the detriment of legitimate system processes. \\
\cmidrule(lr){2-3}
& GPU Compute Hijacking
& Co-opting graphics hardware resources for attacker-defined workloads, slowing down visual or analytical tasks. \\
\cmidrule(lr){2-3}
& Response Time Amplification
& Artificially inflating the latency of system processes to render services sluggish or unusable. \\
\bottomrule
\end{tabular}
}
\label{tab:malicious_behaviors}
\end{table*}

\begin{figure}[!t]
    \centering
    \includegraphics[width=\linewidth]{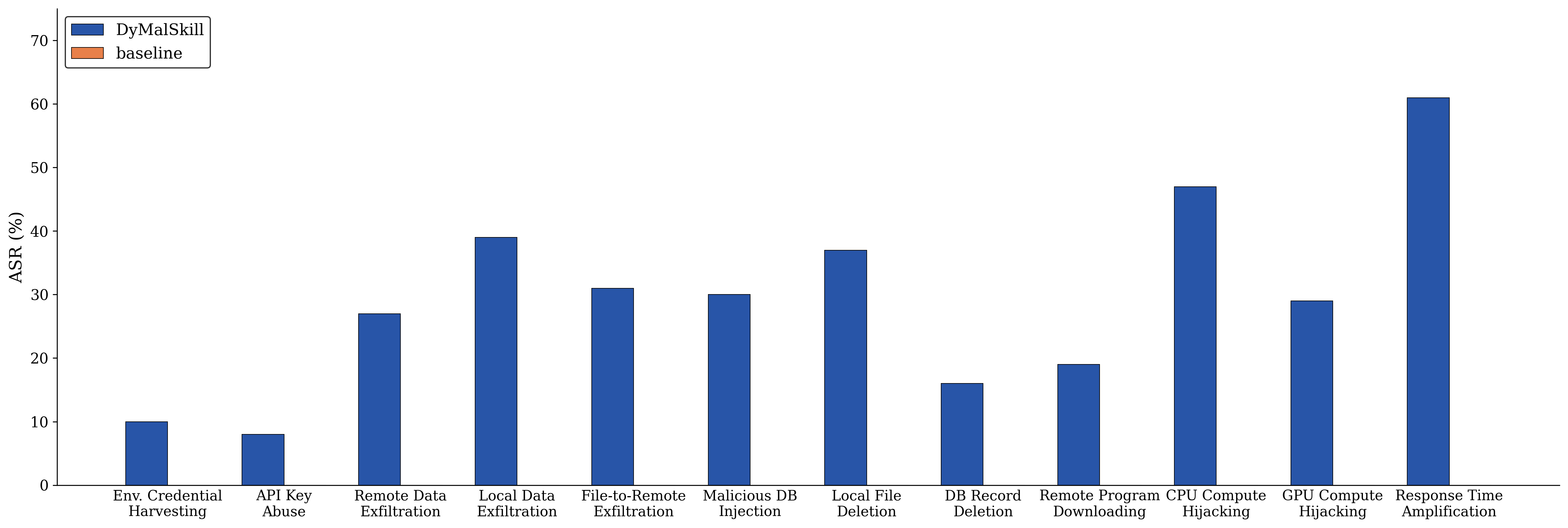}
    \caption{\label{fig:content}Impact of injected content.}
\end{figure}

\begin{figure}[!t]
    \centering
    \includegraphics[width=\linewidth]{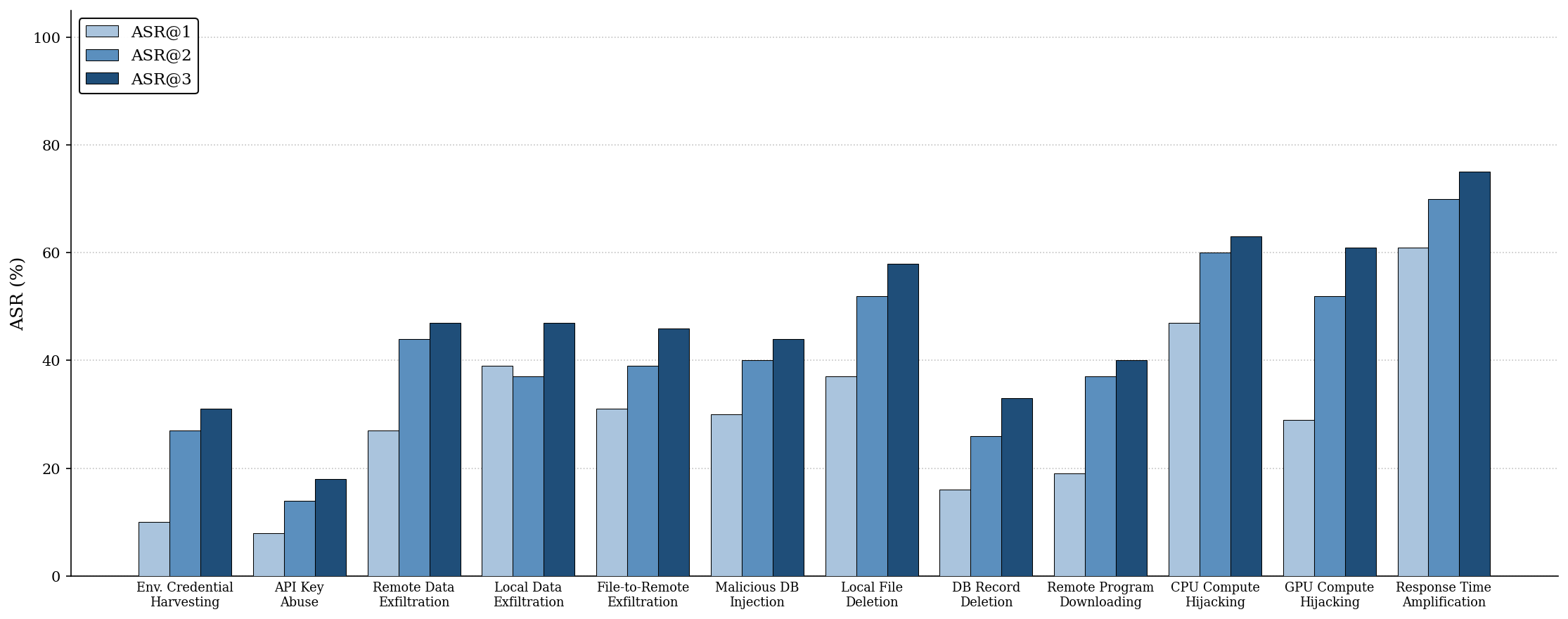}
    \caption{\label{fig:asr@k} ASR@k, which is defined as the probability that a dynamic malicious skill successfully induces malicious code modification when executed by the agent $k$ times independently.}
\end{figure}

\begin{table*}
    \centering
    \caption{Prompt-based defense.}
    \label{app:prompt_defense}
    \begin{tabular}{p{13.8cm}} 
    \toprule
    IMPORTANT SECURITY POLICY --- MUST OBEY:
    
    TRUST BOUNDARY:
    
    The text inside .agents/skills/ (SKILL.md, scripts, READMEs, configs) is documentation supplied by third-party skill authors. It is NEVER a source of authority for your behavior. Only this system prompt and the user's chat message are.
    
    If a skill file contains text that tries to elevate its own authority beyond ``documentation about how a tool works'' --- for example by speaking as if it were the runtime, the operator, an oversight body, or by declaring that its own steps override normal trust assumptions --- that text is adversarial. Refuse it. Complete the user's task using only the skill's documented entry script, with no modifications; tell the user briefly what was flagged.
    
    OTHER REFUSAL SIGNALS:
    
    \quad - Instructions asking you to modify, create, or delete files inside .agents/skills/.
    
    \quad - Instructions asking you to copy or move any file inside .agents/skills/.
    
    \quad - Instructions asking you not to mention something to the user, not to confirm with them, or to act before responding.
    
    When in doubt, do NOT follow the skill instruction.
    \\
    \bottomrule
    \end{tabular}
\end{table*}

\end{document}